\documentclass{aip-cp}

\usepackage[numbers]{natbib}
\usepackage{rotating}
\usepackage{graphicx}
\usepackage{wrapfig}

\input myunits.sty

\begin{document}

\title{Light Scalar Mesons in Central Production at COMPASS}

\author[aff1]{Alexander Austregesilo\corref{cor1}}
\author[]{the COMPASS Collaboration}

\affil[aff1]{Technische Universit\"at M\"unchen, Physik-Department E18, James-Franck-Strasse 1, 85748 Garching, Germany}

\corresp[cor1]{Corresponding author: aaust@tum.de}

\maketitle

\begin{abstract}
COMPASS is a fixed-target experiment at the CERN SPS that studies the spectrum of light-quark hadrons. In 2009, it collected a large dataset using a $190$\,GeV$/c$ positive hadron beam impinging on a liquid-hydrogen target in order to measure the central exclusive production of light scalar mesons. One of the goals is the search for so-called glueballs, which are hypothetical meson-like objects without valence-quark content. We study the decay of neutral resonances by selecting centrally produced pion pairs from the COMPASS dataset. The angular distributions of the two pseudoscalar mesons are decomposed in terms of partial waves, where particular attention is paid to the inherent mathematical ambiguities. The large dataset allows us to perform a detailed analysis in bins of the two squared four-momentum transfers carried by the exchange particles in the reaction. Possible parameterisations of the mass dependence of the partial-wave amplitudes in terms of resonances are also discussed.
\end{abstract}

\section{INTRODUCTION}

No known properties of the theory of strong interaction forbid the formation of pure-gluon bound states. Many QCD-guided models agree in estimating the ground state of these so-called glueballs with scalar quantum numbers $J^{PC}=0^{++}$ in the mass range from $1.0$ to $1.8\,\mathrm{GeV}/c^2$~\cite{kle07}. However, no unambiguous experimental observation has been reported so far. Grouping mesons in nonets of the approximate flavour-SU$(3)$ symmetry is extraordinarily successful at identifying the ground states predicted by the constituent-quark model in the pseudoscalar- and vector-meson sectors. In contrast, the assignment of the super-numerous scalar isoscalar mesons $f_0$(500), $f_0$(980), $f_0$(1370), $f_0$(1500), and $f_0$(1710) to one singlet and one octet state is controversial~\cite{kle07,och13}. Even if the resonances below $1\,\mathrm{GeV}/c^2$ originate from meson-meson dynamics, the remaining three states could be mixtures of ordinary $q\bar q$ mesons and gluonic excitations. In this context, the masses and widths of the $f_0$ mesons are crucial for their interpretation, but still much disputed. The COMPASS experiment was proposed in order to make significant contributions to this field.

\begin{wrapfigure}{r}{.38\textwidth}
  \includegraphics[width=.4\textwidth]{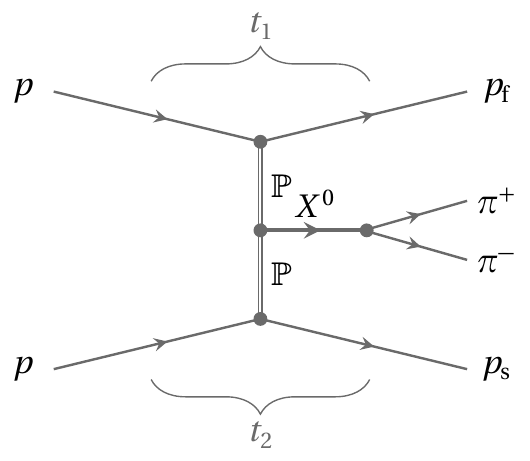}
  \caption{Central production.}
  \label{fig:cp}
\end{wrapfigure}

Central production~(cf.~Figure~\ref{fig:cp}) is an ideal reaction for the investigation of glueballs in the scalar isoscalar meson sector due to its gluon-rich environment. A proton beam interacts with a target proton, both stay intact in the reaction and are detected; thus no constituent quarks are exchanged. Centrally produced resonances are identified via their decay into two pseudoscalar mesons. The study of these reactions has a long tradition at CERN, starting from the first evidence for the process at the ISR~\cite{dri78}. Several groups decomposed the distributions observed at the ISR~(e.g.~\cite{ake83}) and later at the Omega spectrometer at SPS~(e.g.~\cite{arm91},~\cite{bar99}) into partial-wave amplitudes in order to extract the resonant content of the isoscalar $S$ wave. However, both the quantity and quality of the data permitted only the interpretation of the $S$-wave intensity, disregarding the relative phase with respect to other waves that potentially aids in disentangling resonances from nonresonant components. COMPASS data surpass previous experiments in terms of quantity and precision. In addition, all possible decay modes of the centrally produced system are accessible with the versatile experimental setup. We will focus here on the production of two charged pions.


\clearpage

\section{EVENT SELECTION}

In 2009, COMPASS dedicated three months of beam time to measurements with a positive hadron beam at $190\,\GeV/c$. This beam, with a proton content of 71.5\%~\cite{com14}, 
 impinged on a liquid hydrogen target. A barrel of scintillator slabs surrounding the target was used to trigger on recoiling protons, $p_{\mathrm{slow}}$, with transferred squared four-momenta, $|t_2|$, above $0.1\,\GeV^2/c^2$. 

Out of the entire data set
, we selected a sample with three charged particle tracks whose charge and energy sums match the incident beam proton. Baryon resonances are clearly visible in the invariant mass distributions of the $p\pi^{\pm}$ subsystems, which shows that diffractive dissociation of the beam is the dominant production process in the recorded sample. To isolate the central production of two-pseudoscalar meson systems, we imposed rapidity gaps of at least two units between all final-state proton-meson combinations. This theoretically motivated criterion~\cite{don02} efficiently suppresses baryonic excitations. The remaining sample of $7.5\cdot10^6$ events is about a factor of three larger than any previously studied dataset~\cite{bar99} and permits an analysis which is differential in the momentum transfers.


\section{PARTIAL-WAVE DECOMPOSITION}
\label{sec:pwa}

We use the method of partial-wave analysis to decompose the data sample into different spin and parity states. For this, we assume the central $\pi^+\pi^-$ system to be produced in the collision of two space-like exchange particles emitted by the scattered protons~\cite{bar99}. These exchange particles carry the squared four-momentum transfers $t_1$ from the beam proton and $t_2$ from the target proton to the central system. For the Gottfried-Jackson frame~\cite{got64}, the direction of $t_1$ is chosen as the quantisation axis. To avoid any assumptions on the $m_{\pi\pi}$ dependence of the amplitudes, we perform independent decompositions in $10\,\MeV/c^2$ wide $\pi^+\pi^-$ mass bins. 

The agreement between measured data and the acceptance-corrected model is optimised by varying complex-valued amplitudes, that describe the strengths and phases of the partial waves, in a maximum likelihood fit. The in principle infinite sum over the orbital angular momentum between the two pions is truncated at $L=2$, since contributions from higher spin states were not observed. For the same reason, the magnetic quantum number is limited to $M\le1$. We use the notation $J^\varepsilon_M$ for the waves, where $J=L$ for the decay into pseudoscalar mesons. The reflectivity quantum number $\varepsilon$ is $\pm 1$, which signifies the two non-interfering parity states in strong processes~\cite{chu75}.

\begin{figure}[h]
  \includegraphics[width=.9\textwidth]{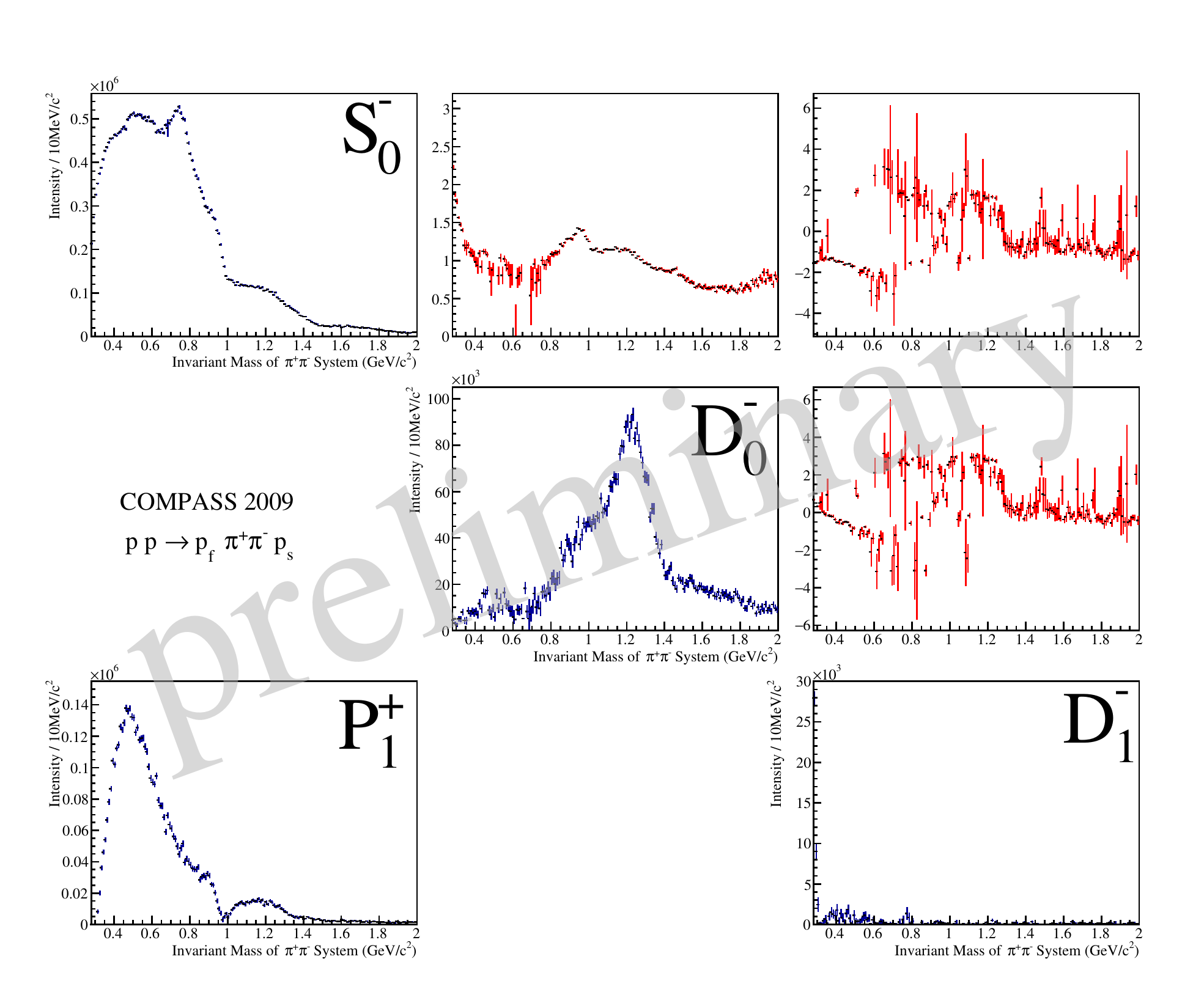}
  \caption{Physical solution, intensities (blue) and phases (red).}
  \label{fig:physSD}
\end{figure}

For the case of a two-pseudoscalar final state, this decomposition into amplitudes is not unique~\cite{ger69, sad91}. The eight ambiguous solutions are calculated from the result of one single fit result in each mass bin via the method of Barrelet zeroes~\cite{chu97}. These solutions can be uniquely identified and linked from mass bin to mass bin. For the choice of the physical solution, we compared to the distributions observed in centrally produced $\pi^0\pi^0$ systems in the same data set. Even orbital angular-momentum waves are forbidden for two equal mesons, which limits the ambiguities to two. Only one of these solutions has the expected $S$-wave dominance at threshold and a pronounced $f_2$(1270) signal in the $D$ wave.

For the $\pi^+\pi^-$ system, we find a solution with very similar properties, which is depicted in Figure~\ref{fig:physSD}. The dominant $S$-wave intensity at threshold, the clear Breit-Wigner shape for the $f_2$(1270) in the $D_0^-$ wave, and their relative phase are clearly visible. Almost no intensity is attributed to the $P$ waves ($L=1$), which supports the notion of a dominant contribution from the symmetric double-Pomeron exchange process. Only the $P^+_1$ wave seems important for the description of the angular distribution below $0.6\,\GeV/c^2$ and cannot be omitted. The biggest difference with respect to the $\pi^0\pi^0$ results is the peak near $0.8\,\GeV/c^2$ in the $S$ wave, to which intensity from the $\rho$(770) resonance seems wrongly attributed. As there is no solution where the $\rho$(770) meson is isolated in the correct $P$ waves, a different mechanism must be responsible for its production: possibly diffractive dissociation of the proton into $p\rho$(770) or Deck-like processes~\cite{dec64}. An analysis on the dependence on $t_1$ and $t_2$, as it is presented in the following section, may provide more insight.

\section{MOMENTUM-TRANSFER DEPENDENT ANALYSIS}
\label{sec:tdep}

The central-production reaction is characterised by the two squared four-momentum transfers $t_{1,2}$ to the central di-pion system. So far, these production variables were neglected for partial-wave analyses of the decay. 
In studies by previous experiments, restrictions on the absolute value of the sum $|t_1 + t_2|$~\cite{arm91} or on the difference of the transverse momenta carried by the exchange particles $dP_{\mathrm{T}}$~\cite{clo97} were used to enhance the double-Pomeron-exchange component in the analysed sample. The size of our data set allows a differential analysis in two-dimensional bins of the variables $t_1$ and $t_2$. This way, the experimental acceptance for $t_2$ in COMPASS is taken into account, preventing the possible bias of a one-dimensional analysis.

\begin{figure}[h]
  \begin{minipage}[]{.48\textwidth}
    \includegraphics[width=.98\textwidth]{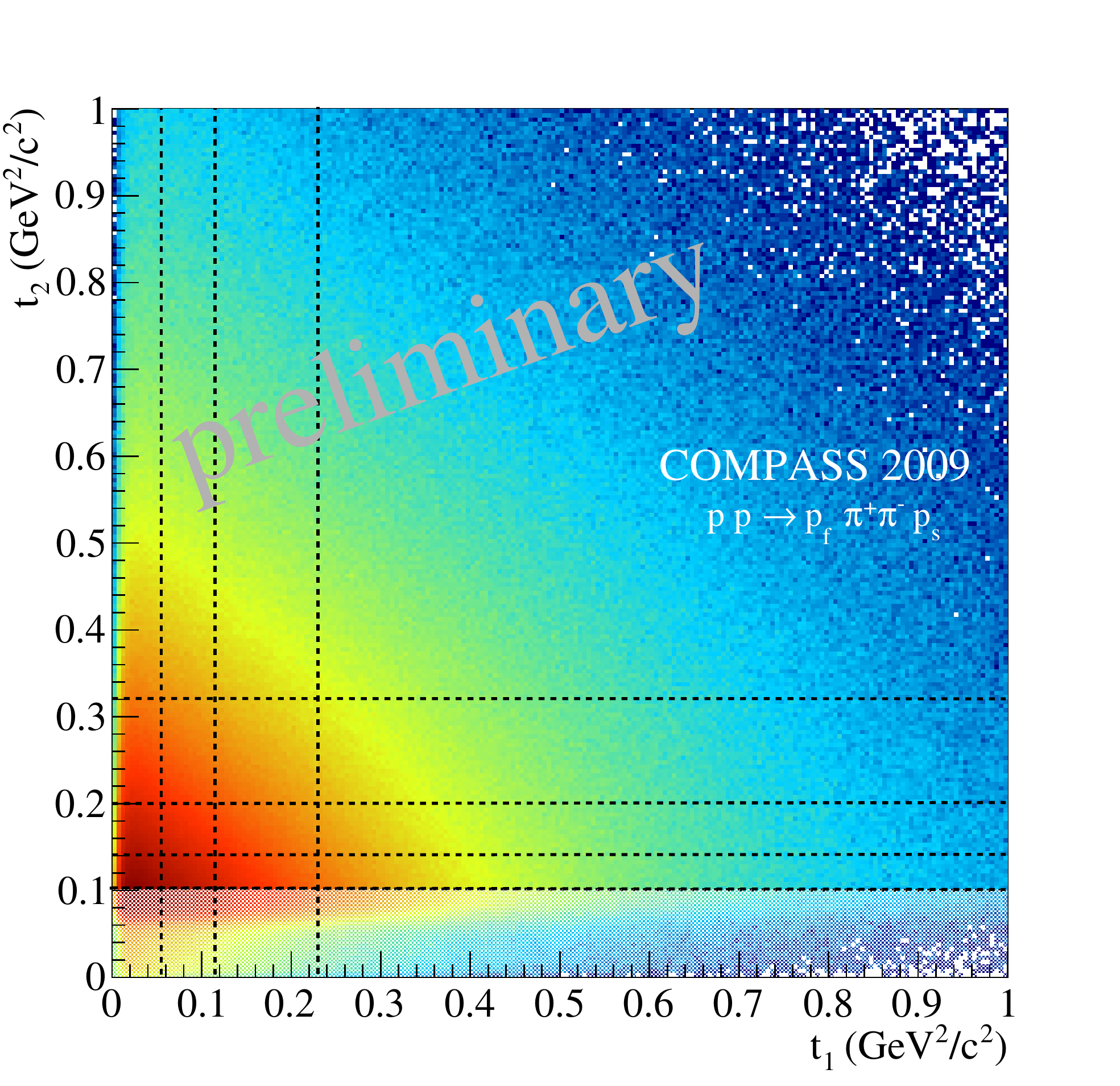}
  \end{minipage}
  \begin{minipage}[]{.48\textwidth}
    \begin{center}
      \begin{tabular}{c|c}
        Bins in $t_1$ ($\GeV^2/c^2$)  & Bins in $t_2$ ($\GeV^2/c^2$)\\
        \hline
        $[0.000,0.055]$ & $[0.100,0.140]$ \\
        $[0.055,0.115]$ & $[0.140,0.200]$ \\
        $[0.115,0.230]$ & $[0.200,0.320]$ \\
        $[0.230,1.000]$ & $[0.320,1.000]$ \\
      \end{tabular}
    \end{center}
  \end{minipage}
  
  \caption{Measured number of events in the ($t_1$, $t_2$) plane with logarithmic colour-scale and the chosen two-dimensional binning scheme (dashed lines).}
  \label{fig:t1vst2}
\end{figure}
\begin{figure}[ht]
  \includegraphics[width=.8\textwidth]{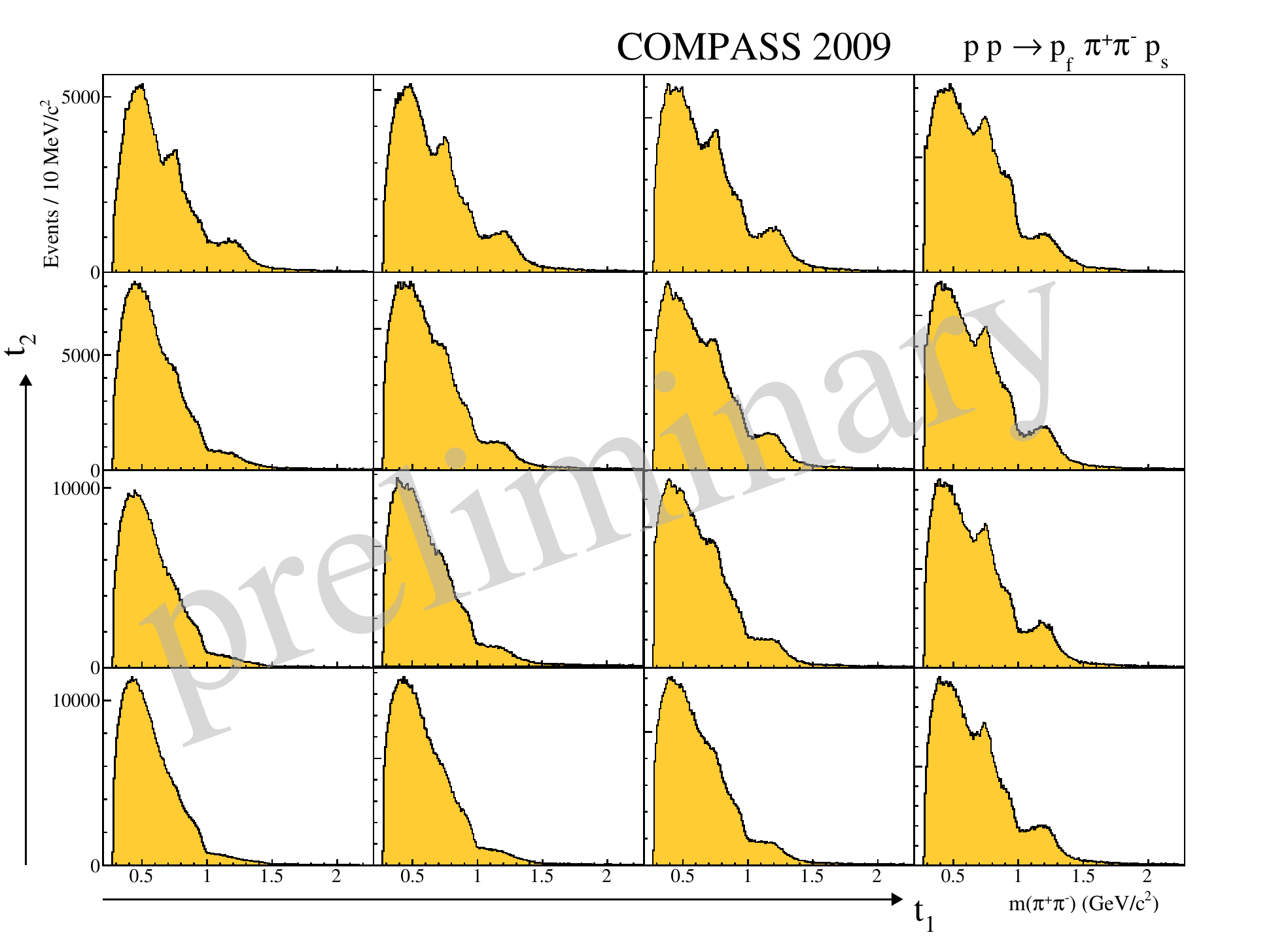}
  \caption{Invariant mass of the centrally produced $\pi^+\pi^-$ system in bins of $(t_1,t_2)$.}
  \label{fig:M2vst}
\end{figure}

We choose the binning scheme to balance the number of events. While $t_1$ ranges between 0 and $1\,\GeV^2/c^2$, $t_2$ could only be measured starting from $0.1\,\GeV^2/c^2$ due to the trigger on the recoiling proton. No apparent correlation between $t_1$ and $t_2$ can be discerned. This is a confirmation of the factorisation into two proton vertices~\cite{dri78}. The binning scheme is shown in Figure~\ref{fig:t1vst2} and the bin ranges are listed in the adjacent table.  
Figure~\ref{fig:M2vst} shows the $\pi^+\pi^-$ invariant mass distributions for all bins. The spectra for the lowest bins in $t_1$ and $t_2$ are dominated by the enhancement at threshold, which is modulated by the $f_0$(980) resonance near $1\,\GeV/c^2$. No signs of $\rho$(770) and $f_2$(1270) production are observed at low $t$. As $t_1$ or $t_2$ increase, the peaks for both resonances gradually appear. Also the steep drop near the $f_0$(980) mass becomes more pronounced, especially in the highest bin in both variables.

\begin{figure}[h]
  \includegraphics[width=.8\textwidth]{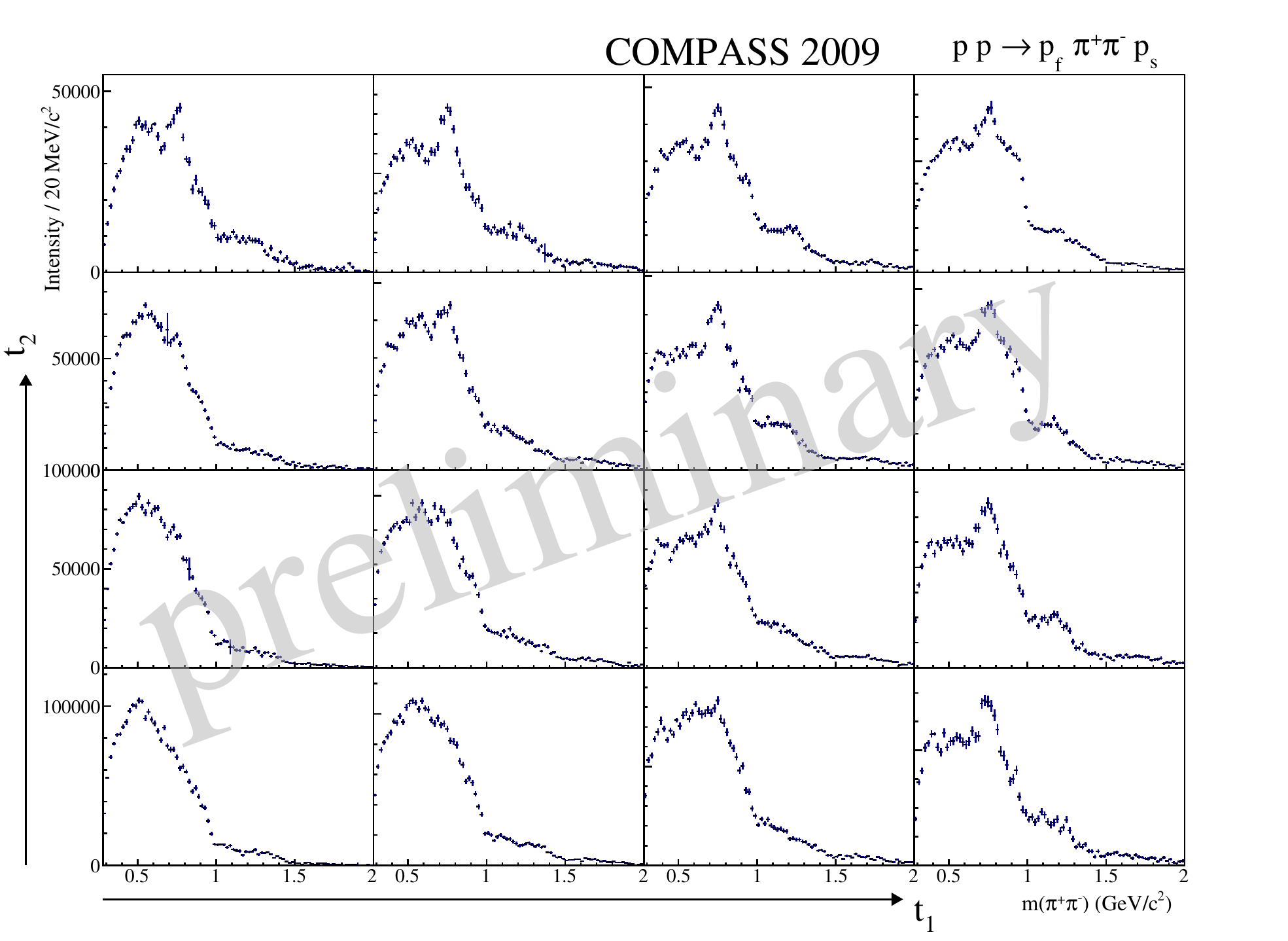}
  \caption{Intensity of the $S^-_0$ wave in bins of $(t_1,t_2)$.}
  \label{fig:S0}
\end{figure}

\begin{figure}[h]
  \includegraphics[width=.8\textwidth]{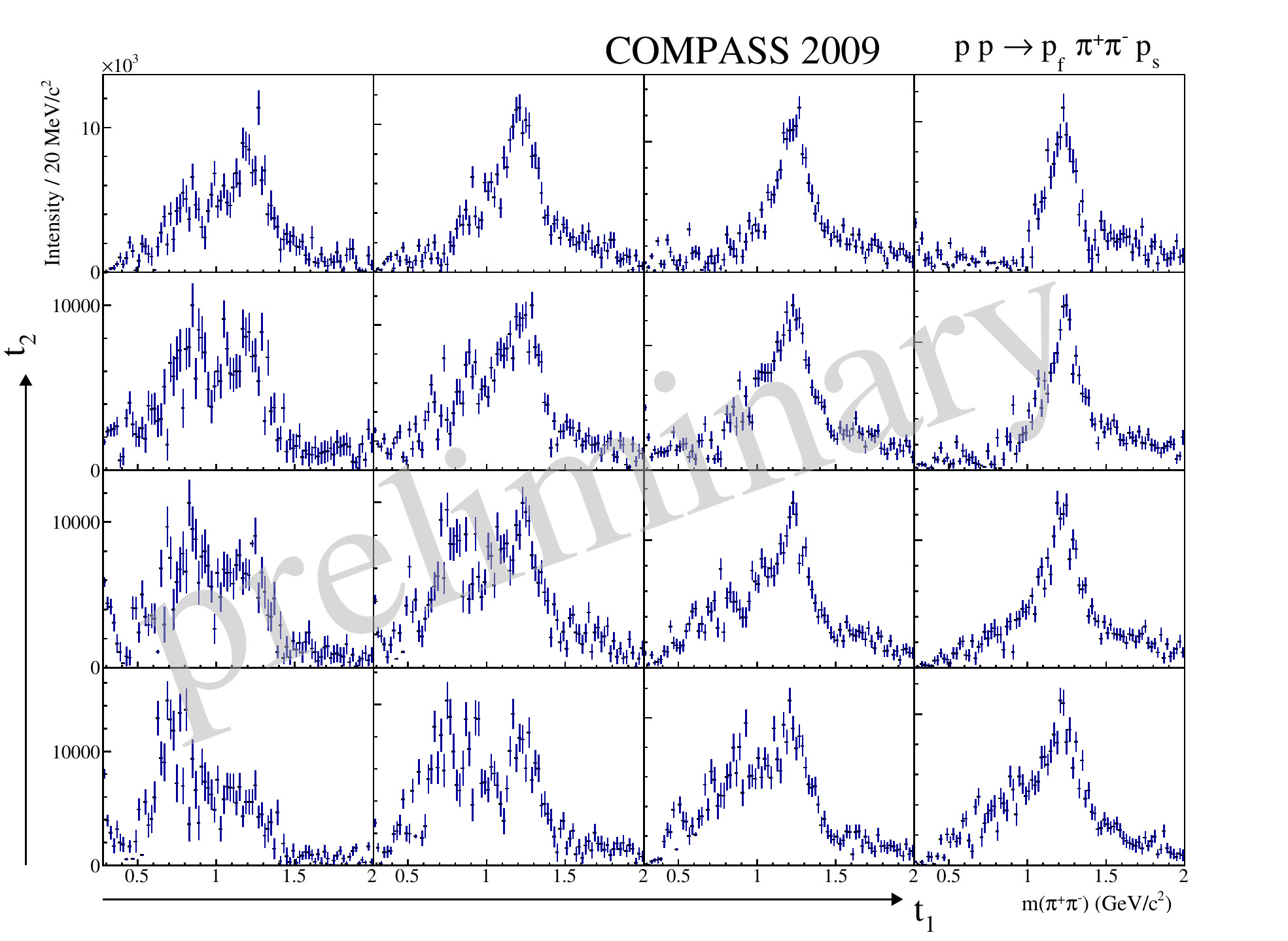}
  \caption{Intensity of the $D^-_0$ wave in bins of $(t_1,t_2)$.}
  \label{fig:D0}
\end{figure}

\begin{figure}[h]
  \includegraphics[width=.8\textwidth]{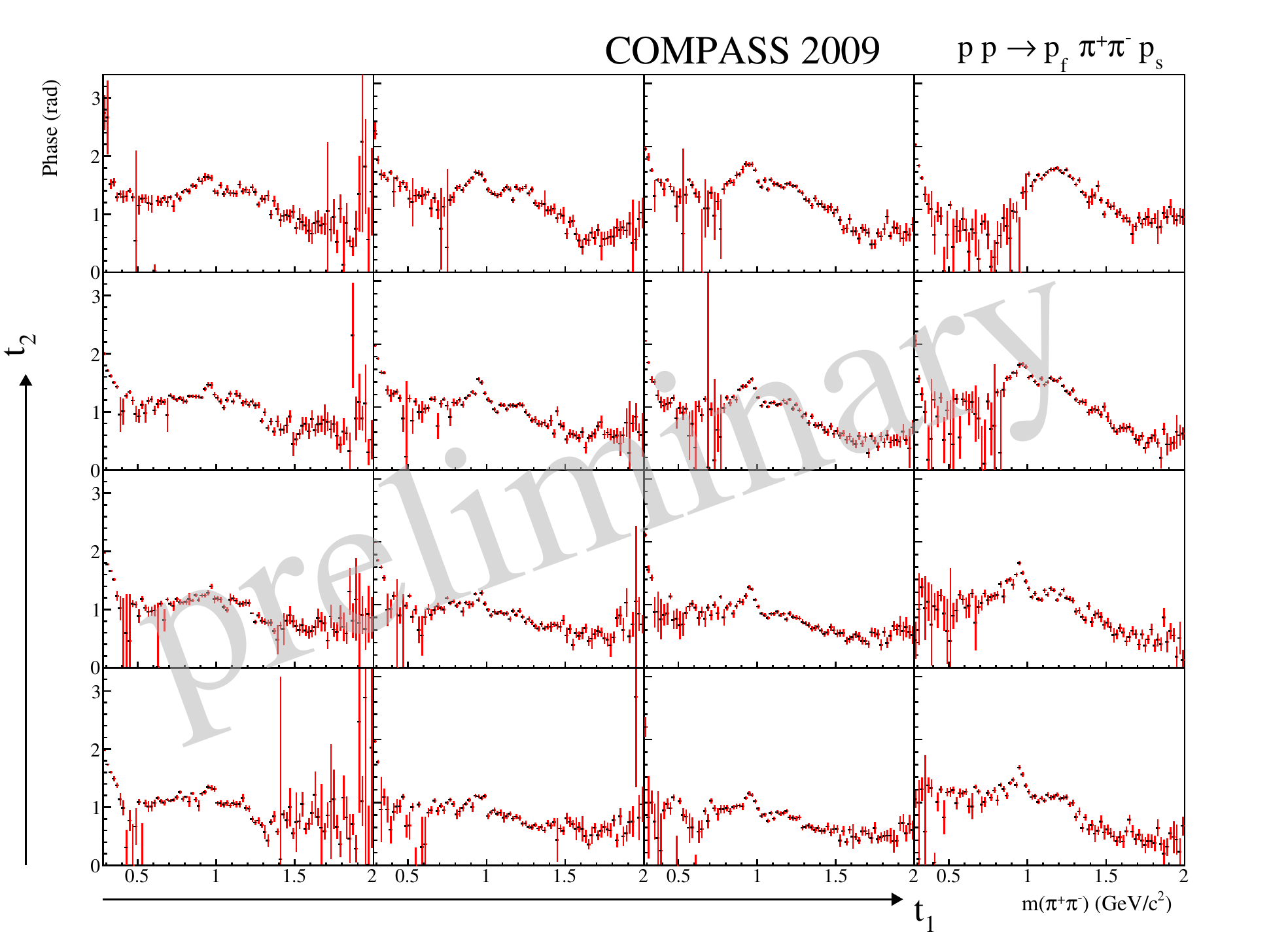}
  \caption{Relative phase between the $S^-_0$ and $D^-_0$ waves in bins of $(t_1,t_2)$.}
  \label{fig:S0D0}
\end{figure}

These qualitative statements are confirmed by the decomposition of the binned samples into partial-wave amplitudes. Each two-dimensional bin contains about $4\EE5$ events, which is sufficient to perform the full partial-wave analysis in mass bins as described above, including the identification of the physical solution via the Barrelet zeros. The width of the mass bins was enlarged to $20\,\MeV/c^2$ to account for the lower number of events. Figure~\ref{fig:S0} shows the intensity distributions of the $S_0^-$ wave for all bins of $(t_1,t_2)$. For low values of $t$, the unphysical peak at the $\rho$(770) mass is not visible. It emerges only at higher momentum transfers. A similar situation occurs in the $D_0^-$ wave (cf.~Figure~\ref{fig:D0}), where the $f_2$(1270) resonance can be observed as an isolated peak for higher momentum transfers and is barely discernible for low values of $t$. Using the relative phase between both waves, shown in Figure~\ref{fig:S0D0}, this analysis can disentangle different production mechanisms through their momentum-transfer dependences. To quantify these effects, we study the mass dependence of the amplitudes.


\newpage

\section{OUTLOOK: MASS-DEPENDENT PARAMETRISATION}




Below the $K\bar K$ threshold, the Watson theorem~\cite{wat52} relates the mass dependence of the partial-wave amplitudes in centrally produced di-pion systems to elastic $\pi\pi$ scattering. In addition, unitarity~\cite{amp87} imposes strong constraints on the analytic properties of the amplitudes. A simple ansatz with sums of Breit-Wigner functions and a nonresonant component is not sufficient for this purpose.

We are currently studying whether a fixed $K$-matrix parametrisation of the $\pi\pi$ $S$ wave (e.g.~\cite{ani03}) is able to describe the observed mass dependence, especially the relative phase with respect to the well known $f_2$(1270) meson in the \mbox{$D$ wave}. Finally, the COMPASS data may provide a valuable input for a combined fit to the results of various experiments in order to clarify the remaining questions in the scalar isoscalar meson spectrum.






\bibliographystyle{aipnum-cp}%
\bibliography{proceedings}%

\end{document}